\documentclass[12pt]{article}
\usepackage{enumerate}
\usepackage{amsfonts}
\usepackage{amsmath}
\usepackage{amssymb}
\usepackage{amsthm}
\usepackage{latexsym}
\usepackage{color}
\usepackage{graphicx}
\usepackage{wrapfig}
\usepackage{caption}
\usepackage{float}

\def\H{\mathcal{H}}

\def\X{\mathcal{X}}

\def\S{\mathfrak{S}}

\def\T{\mathfrak{T}}

\newcommand{\rank}{\mathrm{rank}}

\newcommand{\Tr}{\mathrm{Tr}}

\newcommand{\shs}{\hspace{1pt}}
\newcounter{defin}  \newcounter{lemma}  \newcounter{theorem}
\newcounter{proposition} \newcounter{corol}  \newcounter{remark} \newcounter{example}

\newenvironment{theorem}{\par\refstepcounter{theorem}     \textbf{Theorem \thetheorem.}\ }{\rm\par}
\newenvironment{proposition}{\par\refstepcounter{proposition}     \textbf{Proposition \theproposition.}\ }{\rm\par}
\newenvironment{corollary}{\par\refstepcounter{corol}     \textbf{Corollary \thecorol.} }{\rm\par}

\newenvironment{example}{\par\refstepcounter{example}     \textbf{Example \theexample.}}{\rm\par}

\begin{document}

\title{Upper bounds on the Holevo quantity arising from the fundamental entropic inequality}

\author{M.E.~Shirokov\footnote{email:msh@mi.ras.ru}\\
Steklov Mathematical Institute, Moscow, Russia}
\date{}
\maketitle
\begin{abstract}
We show how the fundamental entropic inequality proved recently in \cite{D&Co} can be applied to obtain
a useful relation for the Holevo quantity of discrete and continuous ensembles of quantum states.

This relation gives a tight upper bound on the Holevo quantity of a given ensemble $\mu$ expressed in terms of  the Holevo quantities
of two auxiliary ensembles $\mu_+$ and $\mu_-$ produced by $\mu$. Among others, this implies quite accurate upper bounds on the Holevo quantity of a discrete ensemble of quantum states expressed via the probabilities and the metric characteristics of an ensemble.
\end{abstract}



\section{Introduction and preliminaries}

The Holevo quantity of an ensemble of quantum states (also called Holevo information or
Holevo bound) is an upper bound on the classical information obtained by recognizing the states of this ensemble by quantum measurements \cite{H-73}. It plays a basic role in analysis of information properties of quantum systems and channels \cite{H-SCI,N&Ch,Wilde}.

The Holevo quantity of a discrete (finite or countable) ensemble\footnote{Ensemble $\{p_{i},\rho_{i}\}_{i\in I}$ is a finite or
countable collection $\{\rho_{i}\}_{i\in I}$ of states
with a probability distribution $\{p_{i}\}_{i\in I}$. The state
$\sum_{i\in I}p_{i}\rho_{i}$ is called \emph{average state} of this  ensemble.} $\mu=\{p_i,\rho_i\}_{i\in I}$ of  quantum states is defined as
$$
\chi\left(\mu\right)\doteq \sum_{i\in I}p_i D(\rho_i\|\shs\bar{\rho}(\mu))=S(\bar{\rho}(\mu))-\sum_{i\in I} p_i S(\rho_i),\quad \bar{\rho}(\mu)=\sum_{i\in I} p_i\rho_i,
$$
where $D(\cdot\|\cdot)$ is the quantum relative entropy, $S(\cdot)$ is the von Neumann entropy (introduced below) and the second formula is valid if $\,S(\bar{\rho}(\mu))<+\infty$. This definition is naturally generalized to continuous ensembles of  quantum states (see Section 3.2).

Anyway, the exact value of the Holevo quantity can be found by calculation of the entropy (relative entropy) for a collection of quantum states, which requires some efforts (especially, in the infinite-dimensional case). Therefore it is convenient to have easy-computable estimates for the Holevo quantity not requiring special calculations.

A problem of finding easy-computable estimates (in particular, upper bounds) for the Holevo quantity was considered by many authors \cite{Aud,B&H,Datta++,Roga, Roga+,LC,HUB}.
In this note we propose and analyse  a general inequality allowing us to construct new upper bounds for the Holevo quantity.
This inequality is proved  by applying the \emph{fundamental entropic inequality}
\begin{equation}\label{b-in}
S(\rho)+\varepsilon S(\tau_{-})\leq S(\sigma)+\varepsilon
S(\tau_{+})+h(\varepsilon),
\end{equation}
valid for any quantum states $\rho$ and $\sigma$ with possible value $+\infty$ in one or both sides, which is recently
proved in \cite{D&Co}. Here $\varepsilon=\frac{1}{2}\|\rho-\sigma\|_1$, $\tau_{+}=(1/\varepsilon)[\rho-\sigma]_+$ and $\tau_i^{-}=(1/\varepsilon)[\rho-\sigma]_-$ are states in $\S(\H)$ ($A_+$ and $A_-$ denote  the positive and negative parts of a Hermitian operator $A$), $\,h(p)=\eta(p)+\eta(1-p)\,$ is the binary entropy ($\eta(x)=x\log(x)$).\bigskip

In Section 2, for a given discrete or continuous ensemble $\mu$ we construct  two auxiliary ensembles $\mu_+$ and $\mu_-$ and use inequality (\ref{b-in}) to show that
\begin{equation}\label{I}
 \chi(\mu)\leq \varepsilon_{av}(\chi(\mu_+)-\chi(\mu_-))+\bar{h}(\mu).
\end{equation}
Here  $\varepsilon_{av}$ is the mean distance from the states of $\mu$ to the average state of $\mu$ and $\bar{h}(\mu)$
is the mean value of the binary entropy on the set $\{\frac{1}{2}\|\rho-\bar{\rho}(\mu)\|_1\}\subset[0,1]$, where $\rho$ runs over the set of states of $\mu$.
\smallskip

In Section 3, we apply  inequality (\ref{I}) to construct upper bounds on the Holevo quantity of discrete ensembles of quantum states expressed via the probabilities and the metric characteristics of these ensembles.\bigskip

Let $\mathcal{H}$ be a separable Hilbert space,
$\mathfrak{B}(\mathcal{H})$ the algebra of all bounded operators on $\mathcal{H}$ with the operator norm $\|\cdot\|$ and $\mathfrak{T}( \mathcal{H})$ the
Banach space of all trace-class
operators on $\mathcal{H}$  with the trace norm $\|\!\cdot\!\|_1$. Let
$\mathfrak{S}(\mathcal{H})$ be  the set of quantum states (positive operators
in $\mathfrak{T}(\mathcal{H})$ with unit trace) \cite{H-SCI,N&Ch,Wilde}.


The \emph{support} $\mathrm{supp}\rho$ of an operator $\rho$ in  $\T_{+}(\H)$ is the closed subspace spanned by the eigenvectors of $\rho$ corresponding to its positive eigenvalues.  The dimension of $\mathrm{supp}\rho$ is called the \emph{rank} of $\rho$ and is denoted by $\rank\rho$.

The \emph{von Neumann entropy} of a quantum state
$\rho \in \mathfrak{S}(\H)$ is  defined by the formula
$S(\rho)=\operatorname{Tr}\eta(\rho)$, where  $\eta(x)=-x\log x$ if $x>0$
and $\eta(0)=0$. It is a concave lower semicontinuous function on the set~$\mathfrak{S}(\H)$ taking values in~$[0,+\infty]$ \cite{H-SCI,O&P,W}.
We will use the inequality
\begin{equation}\label{H-in}
S(\bar{\rho})\leq \sum_{i\in I} p_iS(\rho_i)+H(\{p_i\}_{i\in I}),
\end{equation}
valid for any ensemble  $\{p_i, \rho_i\}_{i\in I}$ of quantum states with the average state $\bar{\rho}$,
where
\begin{equation}\label{SE}
H(\{p_i\}_{i\in I})= \sum_{i\in I}\eta(p_i)
\end{equation}
is the Shannon entropy of the distribution  $\{p_i\}_{i\in I}$ \cite{O&P,N&Ch,Wilde}.\smallskip

\section{The main inequality}

\subsection{Discrete ensembles}

Let $\mu=\{p_i, \rho_i\}_{i\in I}$ be a finite or countable ensemble of quantum states in $\S(\H)$ with the average state
$$
\bar{\rho}(\mu)=\sum_{i\in I}p_i\rho_i.
$$
Introduce two ensembles $\mu_+=\{p_i\varepsilon_i\varepsilon^{-1}_{av}, \tau_i^+\}_{i\in I}$ and $\mu_-=\{p_i\varepsilon_i\varepsilon^{-1}_{av}, \tau_i^-\}_{i\in I}$, where
\begin{itemize}
  \item $\varepsilon_i\doteq\frac{1}{2}\|\rho_i-\bar{\rho}(\mu)\|_1$,
  \item $\varepsilon_{av}\doteq\sum_{i\in I}p_i\varepsilon_i$ -- the mean distance from the states $\{\rho_i\}_{i\in I}$ to the  state $\bar{\rho}(\mu)$,
  \item $\tau_i^{+}=(1/\varepsilon_i)[\rho_i-\bar{\rho}(\mu)]_+$ and $\tau_i^{-}=(1/\varepsilon_i)[\rho_i-\bar{\rho}(\mu)]_-$ are states in $\S(\H)$ for each $i\in I$. \footnote{$[\rho_i-\bar{\rho}(\mu)]_\pm$ are the positive and negative parts of the Hermitian operator $\rho_i-\bar{\rho}(\mu)$.}
\end{itemize}

Concrete examples of ensembles $\mu_+$ and $\mu_-$ can be found in the proof of Theorem \ref{main} and in Example 2 in Section 3.2.\smallskip

 It is essential that the average states $\bar{\rho}(\mu_+)$ and $\bar{\rho}(\mu_-)$ of the ensembles $\mu_+$ and $\mu_-$ coincide:
\begin{equation}\label{omega}
  \bar{\rho}(\mu_+)=\bar{\rho}(\mu_-).
\end{equation}
Indeed,
by definitions we have
$\;
\sum_{i\in I}p_i\varepsilon_i(\tau_i^{-}-\tau_i^{+})=\sum_{i\in I}p_i(\bar{\rho}-\rho_i)=0
$\;
and hence
\begin{equation*}
\frac{1}{\varepsilon_{av}}\sum_{i\in I}p_i\varepsilon_i\tau_i^{+}=\frac{1}{\varepsilon_{av}}\sum_{i\in I}p_i\varepsilon_i\tau_i^{-}.
\end{equation*}

Inequality (\ref{b-in}) allows us to prove the following theorem in which we use the mean value of the binary entropy on the set $\{\frac{1}{2}\|\rho_i-\bar{\rho}(\mu)\|_1\}_{i\in I}\subset[0,1]$ defined as
\begin{equation}\label{h-def}
\bar{h}(\mu)\doteq\sum_{i\in I}p_ih(\varepsilon_i)\qquad \left(\varepsilon_i\doteq\textstyle\frac{1}{2}\|\rho_i-\bar{\rho}(\mu)\|_1\right).
\end{equation}


\begin{theorem}\label{main} \emph{Let $\mu=\{p_i, \rho_i\}_{i\in I}$ be an ensemble of quantum states in $\S(\H)$ with the average state $\bar{\rho}(\mu)$ such that
$S(\bar{\rho}(\mu))<+\infty$. Let $\mu_+$ and $\mu_-$ be the ensembles defined before the theorem. Then 
\begin{equation}\label{main+}
 \chi(\mu)\leq \varepsilon_{av}(\chi(\mu_+)-\chi(\mu_-))+\bar{h}(\mu),
\end{equation}
where $\varepsilon_i\doteq\frac{1}{2}\|\rho_i-\bar{\rho}(\mu)\|_1$,  $\varepsilon_{av}\doteq\sum_{i\in I}p_i\varepsilon_i$ and $\bar{h}(\mu)$ is defined in (\ref{h-def}). The second term in (\ref{main+}) can be replaced by $h(\varepsilon_{av})$.}\smallskip

\emph{The inequality (\ref{main+}) is tight: for any natural $m$ there exists ensemble $\mu$ consisting of $m$ states such that an equality holds in (\ref{main+}). }
\end{theorem}\medskip

\textbf{Note:} The inequality (\ref{main+}) is an improved version of  inequality (7) in \cite{HUB} in the case $\sigma=\bar{\rho}(\mu)$.\smallskip

\emph{Proof.} We will denote the state $\bar{\rho}(\mu)$ by $\bar{\rho}$ for brevity.  Since $S(\bar{\rho})<+\infty$, we have
$$
\chi(\mu)=S(\bar{\rho})-\sum_{i\in I}p_iS(\rho_i)=\sum_{i\in I}p_i(S(\bar{\rho})-S(\rho_i)).
$$
By inequality (\ref{b-in}) we have
\begin{equation}\label{one}
\chi(\mu)=\sum_{i\in I}p_i(S(\bar{\rho})-S(\rho_i))\leq\sum_{i\in I}p_i\varepsilon_i(S(\tau_i^{-})-S(\tau_i^{+}))+\sum_{i\in I}p_ih(\varepsilon_i)
\end{equation}
By using (\ref{omega}) and by noting that the state $\bar{\rho}(\mu_+)=\bar{\rho}(\mu_-)$ has finite entropy (because $S(\bar{\rho})<+\infty$), we obtain
$$
\varepsilon_{av}\sum_{i\in I}p_i\varepsilon_i\varepsilon^{-1}_{av}(S(\tau_i^{-})-S(\tau_i^{+}))=\varepsilon_{av}(\chi(\mu_+)-\chi(\mu_-))
$$
By inserting this into the inequality (\ref{one}) we obtain  (\ref{main+}).\smallskip

The last claim of the main part of the theorem  follows from the concavity of the binary entropy.\smallskip

To prove the tightness of (\ref{main+}) take natural $m$ and consider
an ensemble $\mu$ of $m$ mutually orthogonal pure states with equal probabilities. In this case
$$
\chi(\mu)=\log m,\quad  \varepsilon_{i}=\varepsilon_{av}=1-1/m \quad i=\overline{1,m}.
$$
It is easy to see that $\mu_+=\mu$, while $\mu_-=\{q_i,\sigma_i\}_{i=1}^m$, where $\sigma_i=\frac{1}{m-1}(m\bar{\rho}(\mu)-\rho_i)$.
So, in this case $\chi(\mu_+)=\log m$, $\chi(\mu_-)=\log m-\log (m-1)$ and hence
$$
 \chi(\mu)=(1-1/m)(\chi(\mu_+)-\chi(\mu_-))+h(1-1/m).
$$
This equality means the equality case in (\ref{main+}). $\Box$

\subsection{Continuous ensembles}

In this subsection we generalize the results of the previous one to "continuous" ensembles of quantum states. 

Let $\X$ be a measurable space. According to the commonly used notation (cf.\cite{H-SCI,H-GE}) a generalized ensemble $\{\pi, \rho_x\}_{x\in\X}$
consists of a probability measure $\pi$  on $\X$ and a "measurable family" $\{\rho_x\}_{x\in\X}$ of
quantum states on $\H$. It means, mathematically, that $\rho_x=\Upsilon(x)$, where $\Upsilon$ is a measurable $\S(\H)$-valued function on $\X$.
In fact, any such ensemble can be treated as a Borel probability measure on $\S(\H)$ corresponding to the image $\Upsilon(\pi)$ of the measure $\pi$
under the map $\Upsilon$. So, the above definition of a continuous ensemble agrees with the notion of a generalized ensemble proposed in \cite{H-Sh-2}.

For a given continuous ensemble $\mu=\{\pi, \rho_x\}_{x\in\X}$  with the average state 
$$
\bar{\rho}(\mu)\doteq \int_{\X} \rho_x \pi(dx)
$$
the Holevo quantity is defined as
$$
\chi(\mu)\doteq \int_{\X}D(\rho_x\,\|\bar{\rho}(\mu))\pi(dx)=S(\bar{\rho}(\mu))-\int_{\X}S(\rho_x)\pi(dx),
$$
where the second formula is valid if $\,S(\bar{\rho}(\mu))<+\infty\,$ \cite{H-SCI,H-Sh-2}.\smallskip

Introduce 
two continuous ensembles $\mu_+=\{\pi_{\pm}, \tau^+_x\}_{x\in\X}$ and $\mu_-=\{\pi_{\pm}, \tau^-_x\}_{x\in\X}$, 
where
\begin{itemize}
  \item $\varepsilon_x=\frac{1}{2}\|\rho_x-\bar{\rho}(\mu)\|_1$ is a measurable function on $\X$,
  \item $\varepsilon_{av}\doteq\int_{\X}\varepsilon_x\pi(dx)$ -- the mean distance from the states $\{\rho_x\}_{x\in \X}$ to the  state $\bar{\rho}(\mu)$,
  \item $\tau_x^{+}=(1/\varepsilon_x)[\rho_x-\bar{\rho}(\mu)]_+$ and $\tau_x^{-}=(1/\varepsilon_x)[\rho_x-\bar{\rho}(\mu)]_-$ are states in $\S(\H)$ for each $x\in \X$ (if $\varepsilon_x=0$ then $\tau_x^{+}=\tau_x^{-}=\sigma_*$ -- a fixed arbitrary state in $\S(\H)$).\footnote{$[\rho_x-\bar{\rho}(\mu)]_\pm$ are the positive and negative parts of the Hermitian operator $\rho_x-\bar{\rho}(\mu)$.}
  \item $\pi_{\pm}$ is the probability measure on $\X$, which has the density $\varepsilon^{-1}_{av}\varepsilon_x$ w.r.t. the measure $\pi$.
\end{itemize}

By the obvious modification of the arguments from Section 3.1 we conclude that 
\begin{equation}\label{omega+}
  \bar{\rho}(\mu_+)=\bar{\rho}(\mu_-).
\end{equation}

To formulate  a generalized (continuous) version of Theorem \ref{main} we have to define the mean value of the binary entropy on the set $\{\frac{1}{2}\|\rho_x-\bar{\rho}(\mu)\|_1\}_{x\in \X}\subset[0,1]$ as
\begin{equation}\label{h-def+}
\bar{h}(\mu)\doteq\int_{\X}h(\varepsilon_x)\pi(dx)\qquad \left(\varepsilon_x\doteq\textstyle\frac{1}{2}\|\rho_x-\bar{\rho}(\mu)\|_1\right).
\end{equation}


\begin{theorem}\label{main-g} \emph{Let $\X$ be a metric space and $\rho_x$ a continuous $\S(\H)$-valued function on $\X$. Let $\mu=\{\pi, \rho_x\}_{x\in\X}$ be a generalized  ensemble such that $S(\bar{\rho}(\mu))<+\infty$. Let $\mu_+$ and $\mu_-$ be the ensembles defined before the theorem. Then 
\begin{equation}\label{main-g+}
 \chi(\mu)\leq \varepsilon_{av}(\chi(\mu_+)-\chi(\mu_-))+\bar{h}(\mu),
\end{equation}
where  $\varepsilon_{av}$ is the parameter defined before the theorem and $\bar{h}(\mu)$ is defined in (\ref{h-def+}). The second term in (\ref{main-g+}) can be replaced by $h(\varepsilon_{av})$.}\smallskip

\emph{The inequality (\ref{main-g+}) is tight:  there are ensembles $\mu$ for which  an equality holds in (\ref{main-g+}).}
\end{theorem}\medskip

\emph{Proof.} The claims of this  theorem are deduced from the corresponding claims of Theorem \ref{main} by applying the  approximation technique 
used in  the proof Proposition 9 in \cite{NMP}. $\Box$

\section{Applications}

\subsection{Upper bound on $\chi(\{p_i,\!\rho_i\})$ depending on $\{p_i\}$ and $\{\|\rho_i\!-\!\bar{\rho}\|_1\}$}

Since  the Holevo quantity $\chi(\{q_i, \sigma_i\}_{i\in I})$ of any ensemble is nonnegative and bounded from above
by the Shannon entropy $H(\{q_i\}_{i\in I})$ of the probability distribution $\{q_i\}_{i\in I}$, Theorem \ref{main} implies the following
\smallskip

\begin{proposition}\label{main_b} \emph{Let $\mu=\{p_i, \rho_i\}_{i\in I}$ be an ensemble of quantum states in $\S(\H)$ with the average state $\bar{\rho}(\mu)$ such that  either $S(\bar{\rho}(\mu))<+\infty$ or $H(\{p_i\}_{i\in I}\})<+\infty$. Then 
\begin{equation}\label{main_b+}
 \chi(\mu)\leq \varepsilon_{av}H\left(\left\{\frac{p_i\varepsilon_i}{\varepsilon_{av}}\right\}_{i\in I}\right)+\sum_{i\in I}p_ih(\varepsilon_i),
\end{equation}
where $\varepsilon_i\doteq\frac{1}{2}\|\rho_i-\bar{\rho}(\mu)\|_1$,  $\varepsilon_{av}\doteq\sum_{i\in I}p_i\varepsilon_i$ -- the mean  distance from the states $\{\rho_i\}_{i\in I}$ to the average state $\bar{\rho}(\mu)$, $H(\cdot)$ is the Shannon entropy and $h(\cdot)$ is the binary entropy}.\smallskip

\emph{The second term in (\ref{main_b+}) can be replaced by $h(\varepsilon_{av})$. }
\end{proposition}\smallskip

\textbf{Note:} The  second term in the r.h.s. of (\ref{main_b+}) cannot be omitted. To see this it suffices to consider
an ensemble of $d$ mutually orthogonal pure states, since in this case
$$
\chi(\{p_i, \rho_i\})=\log d,\quad \textrm{while } \quad \varepsilon_{av}H\left(\left\{\frac{p_i\varepsilon_i}{\varepsilon_{av}}\right\}_{i\in I}\right)=(1-1/d)\log d.
$$
For this ensemble the difference between  the r.h.s. and the l.h.s. of (\ref{main_b+}) is equal to
$$
-(1-1/d)\log(1-1/d)=o(1)\quad \textrm{as}\quad d\to+\infty. 
$$

\emph{Proof.} We may assume that $p_i>0$ for all $i\in I$ and that $\varepsilon_{av}>0$ (if $\varepsilon_{av}=0$ then both sides of (\ref{main_b+}) are equal to zero). We will denote the state $\bar{\rho}(\mu)$ by  $\bar{\rho}$ for brevity.
\smallskip

If $S(\bar{\rho})<+\infty$  then (\ref{main_b+})  follows directly from Theorem \ref{main} and the remark before the proposition.

Assume that $H(\{p_i\}_{i\in I}\})<+\infty$.  Let $\{P_n\}$ be a nondecreasing sequence of finite rank projectors strongly converging to the unit operator $I_{\H}$.
We may assume that $c_n\doteq\Tr P_n\bar{\rho}>0$ for all $n$.

For each $n$ consider the ensemble $\{p^n_i, \rho^n_i\}_{i\in I}$, where
$$
p_i^n=\frac{c_i^np_i}{c_n},\quad c_i^n=\Tr P_n\rho_i,\quad\textrm{ and }\quad \rho^n_i=\left\{\begin{array}{ll}
        (1/c_i^n)P_n\rho_iP_n\; &\textrm{if}\;\;  c_i^n\neq0\\
        \rho_i \;\;&\textrm{otherwise}
        \end{array}\right.
$$

Since the average state $\,\bar{\rho}_n=(1/c_n)P_n \bar{\rho}P_n\,$ of this ensemble has finite rank, the claim of the proposition concerning the case $S(\bar{\rho}(\mu))<+\infty$ proved before implies
\begin{equation}\label{main+n}
 \chi(\{p^n_i, \rho^n_i\})\leq \varepsilon^n_{av}H\left(\left\{\frac{p_i^n\varepsilon^n_i}{\varepsilon^n_{av}}\right\}_{i\in I}\right)+\sum_{i\in I}p^n_ih(\varepsilon^n_i),
\end{equation}
where $\varepsilon^n_i\doteq\frac{1}{2}\|\rho^n_i-\bar{\rho}_n\|_1$ and $\varepsilon^n_{av}\doteq\sum_{i\in I}p^n_i\varepsilon^n_i$.
Since
\begin{equation}\label{lim}
\lim_{n\to+\infty}p_i^n=p_i\quad\textrm{ and }\quad\lim_{n\to+\infty}\rho_i^n=\rho_i\quad \forall i\in I,
\end{equation}
the lower semicontinuity of the relative entropy implies
$$
\chi(\{p_i, \rho_i\})\leq \liminf_{n\to+\infty} \chi(\{p^n_i, \rho^n_i\}).
$$
Thus, to complete this step it suffices to show that the r.h.s. of (\ref{main+n})
tends to the r.h.s. of (\ref{main_b+}) as $n\to+\infty$. As $h(x)\leq \log 2$ for all $x\in[0,1]$,
the first limit relation in (\ref{lim}) implies that
$$
\lim_{n\to+\infty}\sum_{i\in I}p^n_ih(\varepsilon^n_i)=\sum_{i\in I}p_ih(\varepsilon_i),
$$
because
\begin{equation}\label{eps}
 \lim_{n\to+\infty}\varepsilon_i^n=\varepsilon_i\quad \forall i\in I.
\end{equation}

The first limit relation in (\ref{lim}) and (\ref{eps}) show that
$$
\lim_{n\to+\infty}\varepsilon_{av}^n=\varepsilon_{av}>0
$$
and that
\begin{equation}\label{lim+}
\ell_1\textup{-}\lim_{n\to+\infty}\{p^n_i\}_{i\in I}=\{p_i\}_{i\in I},\quad \quad \ell_1\textup{-}\lim_{n\to+\infty}\{\varepsilon_i^n p^n_i\}_{i\in I}=\{\varepsilon_ip_i\}_{i\in I}
\end{equation}
(by the classical version of the well known result in \cite{D-A}).
Hence, $\varepsilon_{av}^n\geq\varepsilon_{av}/\sqrt{2}$ for all sufficiently large $n$.
It follows that
$$
\frac{p_i^n\varepsilon^n_i}{\varepsilon^n_{av}}\leq\frac{\sqrt{2}p_i}{\varepsilon^n_{av}}\leq\frac{2p_i}{\varepsilon_{av}}\quad \forall i\in I
$$
for all sufficiently large $n$, where the first inequality is due to  the first limit relation in (\ref{lim+}).  Thus, as $H(\{p_i\}_{i\in I}\})<+\infty$, the classical version of Simon's  dominated convergence theorem for the entropy (cf.\cite[the Appendix]{Ruskai}) along with the second limit relation in (\ref{lim+})  show that
$$
\lim_{n\to+\infty}H\left(\left\{\frac{p_i^n\varepsilon^n_i}{\varepsilon^n_{av}}\right\}_{i\in I}\right)=H\left(\left\{\frac{p_i\varepsilon_i}{\varepsilon_{av}}\right\}_{i\in I}\right).
$$

The last claim of the proposition follows from the concavity of the binary entropy. $\Box$
\medskip

\begin{example}\label{main-r} As an example allowing us to estimate the accuracy of the upper bound (\ref{main_b+})
consider the ensemble $\mu(N)=\{(1-q_N)q_N^n, |n\rangle\langle  n|\}_{n=0}^{+\infty}$, $q_N=\frac{N}{N+1}$, of pure states of a quantum oscillator,
where $\{|n\rangle\}_{n=0}^{+\infty}$ is the Fock basis in $\H$. The average state of this ensemble is the Gibbs state $\gamma(N)$ corresponding to the mean number of quanta $N$ and, hence, the Holevo quantity $\chi(\mu(N))$ is equal to $g(N)=(N+1)\log(N+1)-N\log(N)$. Since $\frac{1}{2}\|\gamma(N)-|n\rangle\langle  n|\|_1=1-(1-q_N)q_N^n$, the upper bound (\ref{main_b+})
corresponding to the above ensemble is equal to
$$
\widehat{\chi}(\mu(N))=a(N)H\left(\left\{\frac{\lambda_n(1-\lambda_n)}{a(N)}\right\}_{n=0}^{+\infty}\right)+\sum_{n=0}^{+\infty}\lambda_n h(1-\lambda_n),\quad a(N)=\frac{2q_N}{1+q_N}.
$$
Both quantities $\chi(\mu(N))$ and  $\widehat{\chi}(\mu(N))$ are easily calculated, the results  are shown on Figure 1. We see that the difference
is relatively small for different values of $N$.
\end{example}\medskip

\begin{figure}[t]

\centering

\begin{center}

\includegraphics[scale=0.4, bb=400  450 500 550]{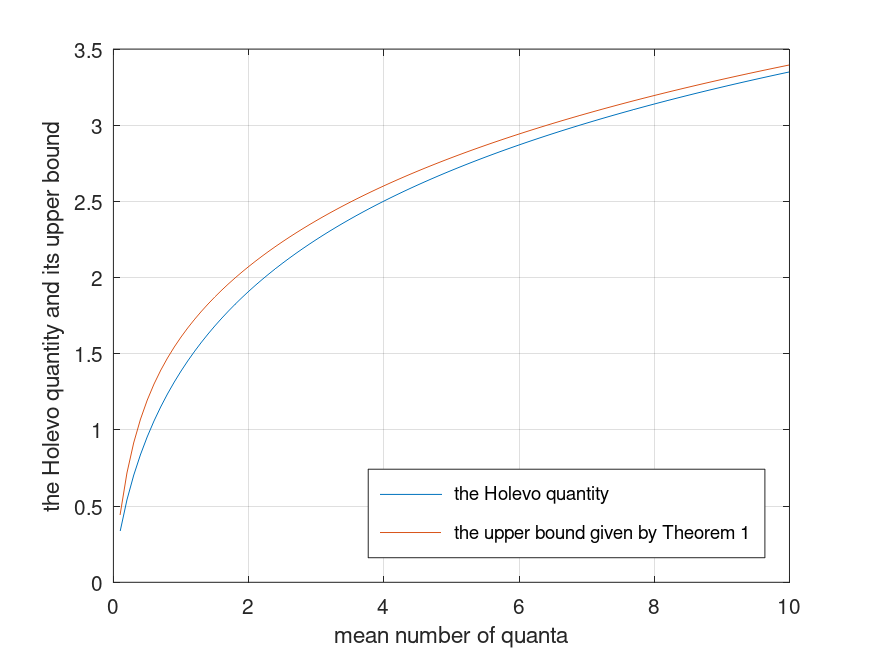}

\vspace{180pt}
\caption{The values of $\chi(\mu(N))$ and $\widehat{\chi}(\mu(N))$ as functions of $N$}
\end{center}

\label{Fig1}
\end{figure}

\begin{corollary}\label{main-c}
\emph{Let $\{p_i, \rho_i\}_{i=1}^m$ be an ensemble of $m<+\infty$ quantum states in $\S(\H)$. Then
\begin{equation}\label{main-c+}
 \chi(\{p_i, \rho_i\}_{i=1}^m)\leq \varepsilon_{av}\log m+\sum_{i=1}^m p_ih(\varepsilon_i)\leq \varepsilon_{av}\log m+h(\varepsilon_{av}),
\end{equation}
where $\varepsilon_i\doteq\frac{1}{2}\|\rho_i-\bar{\rho}(\mu)\|_1$ and $\varepsilon_{av}\doteq\sum_{i=1}^m p_i\varepsilon_i$ -- the mean distance from the states $\{\rho_i\}_{i=1}^m$ to the average state $\bar{\rho}(\mu)$}.
\end{corollary}\medskip

Inequality (\ref{main-c+}) refines the upper bounds for the Holevo quantity depending on $\varepsilon_{av}$ and the number $m$ of ensemble states  obtained in \cite{HUB}.

\subsection{On improvements of the  bound in Proposition \ref{main_b} using information about the operators $[\rho_i-\bar{\rho}(\mu)]_{\pm}$}

The upper bound on the Holevo quantity presented in Proposition \ref{main_b} can be improved by optimizing the proof of this proposition  with the use of information about the operators $[\rho_i-\bar{\rho}(\mu)]_{\pm}$, $i\in I$ ($[\rho_i-\bar{\rho}(\mu)]_{+}$ and $[\rho_i-\bar{\rho}(\mu)]_{-}$ are the positive and negative parts of the operator $\rho_i-\bar{\rho}(\mu)$). \smallskip

\begin{proposition}\label{main++} \emph{Let $\mu=\{p_i, \rho_i\}_{i\in I}$ be an ensemble of quantum states in $\S(\H)$ with the average state $\bar{\rho}(\mu)$ such that either $S(\bar{\rho}(\mu))<+\infty$ or $H(\{p_i\}_{i\in I}\})<+\infty$. Then 
\begin{equation}\label{main+++}
 \chi(\mu)\leq \varepsilon_{av}C(\mu)H\left(\left\{\frac{p_i\varepsilon_i}{\varepsilon_{av}}\right\}_{i\in I}\right)+\sum_{i\in I}p_ih(\varepsilon_i)-\varepsilon_{av}D(\mu),
\end{equation}
where $\varepsilon_i\doteq\frac{1}{2}\|\rho_i-\bar{\rho}(\mu)\|_1$,  $\varepsilon_{av}\doteq\sum_{i\in I}p_i\varepsilon_i$, $H(\cdot)$ is the Shannon entropy, $h(\cdot)$ is the binary entropy and}
$$
C(\mu)=\frac{1}{2}\sup_{i,j\in I}\|\varepsilon^{-1}_i[\rho_i-\bar{\rho}(\mu)]_{+}-\varepsilon^{-1}_j[\rho_j-\bar{\rho}(\mu)]_{+}\|_1
$$
$$
D(\mu)=\frac{\log e}{2}\sum_{i\in I}p_i\|\varepsilon^{-1}_i[\rho_i-\bar{\rho}(\mu)]_{-}-\bar{\rho}(\mu_-)\|^2_1
$$
$$
\left(\bar{\rho}(\mu_-)\doteq\varepsilon^{-1}_{av}\sum_{i\in I}p_i[\rho_i-\bar{\rho}(\mu)]_{-}=\bar{\rho}(\mu_+)\doteq\varepsilon^{-1}_{av}\sum_{i\in I}p_i[\rho_i-\bar{\rho}(\mu)]_{+}\right),
$$
\smallskip

\emph{The second term in (\ref{main+++}) can be replaced by $h(\varepsilon_{av})$. }
\end{proposition}\medskip


\emph{Proof.} It suffices to upgrade the proof of Proposition \ref{main_b} (based on inequality (\ref{main+})) by noting that
\begin{equation}\label{one-in+}
\chi(\mu_{-})\geq D(\mu)
\end{equation}
and that
\begin{equation}\label{two-in+}
\chi(\mu_{+})\leq C(\mu)H\left(\left\{\frac{p_i\varepsilon_i}{\varepsilon_{av}}\right\}_{i\in I}\right),
\end{equation}
where $\mu_{-}$ and $\mu_{+}$ are the ensembles introduced before Theorem \ref{main}.
The inequality (\ref{one-in+})  is a corollary of the well-known lower bound for the quantum relative entropy \cite{H-SCI,Wilde}.
The inequality (\ref{two-in+})  follows from Audenaert's  upper bound on the Holevo quantity \cite{Aud}. $\Box$

\begin{example}\label{two} Consider the ensemble $\hat{\mu}=\{p_i,\rho_i\}_{i=1}^3$ in $\S(\H_2)$, where
$p_1=p_2=p_3=1/3$ and $\rho_1$, $\rho_2$, $\rho_3$ are pure states corresponding to the vectors
$$
\left[\begin{array}{c}
  1 \\
  0
\end{array}\right],\qquad \left[\begin{array}{c}
  -1/2 \\
  \sqrt{3}/2
\end{array}\right]\qquad \left[\begin{array}{c}
  -1/2 \\
  -\sqrt{3}/2\end{array}\right],
$$
Then $\bar{\rho}(\hat{\mu})=I_2/2$ and  $\varepsilon_1=\varepsilon_2=\varepsilon_3=1/2$. Hence, in this case
the upper bound (\ref{main_b+}) is equal to $(1/2)\log 3+\log 2\approx1.24$ (we assume that $\log=\ln$),
which is essentially greater than $\chi(\hat{\mu})=\log2\approx 0.69$.

Since $\hat{\mu}_+=\hat{\mu}$ and $\hat{\mu}_-=\{q_i,\sigma_i\}_{i=1}^3$, where
$q_1=q_2=q_3=1/3$ and $\sigma_1=I_2-\rho_1$, $\sigma_2=I_2-\rho_2$, $\sigma_3=I_2-\rho_3$, we have
$$
C(\hat{\mu})=\sqrt{1-1/4}=\sqrt{3}/2\quad\textrm{ and }\quad D(\hat{\mu})=1/2.
$$
So, the upper bound (\ref{main+++}) gives more accurate estimate $(\sqrt{3}/4)\log 3+\log 2-1/4\approx 0.92$.\smallskip

Note, finally, that inequality  (\ref{main+}) provides a sharp upper bound on $\chi(\hat{\mu})$. Indeed,  since $\chi(\hat{\mu}_+)=\chi(\hat{\mu}_-)=\log 2$,
we have
$$
\chi(\hat{\mu})=\log2=(1/2)(\chi(\hat{\mu}_+)-\chi(\hat{\mu}_-))+h(1/2).
$$
\end{example}

\bigskip 
I am grateful to  Nilanjana Datta and Shang Cheng for useful comments and references.

\end{document}